\documentstyle[prd,aps]{revtex}
\begin{document}
\draft
\baselineskip 16pt
\renewcommand{\topfraction}{0.8}
\renewcommand{\textwidth}{15cm}
\renewcommand{\topmargin}{-0.5cm}
\preprint{OU-TAP }
%\date{February, 2001}
\tighten
\title{After bulky brane inflation %
}

\author{ Jun'ichi YOKOYAMA and Yoshiaki HIMEMOTO}
\address{\hfill\\
Department of Earth and Space Science, Graduate School of Science,\\
Osaka University, Toyonaka 560-0043, Japan\\
%\hfill\\
}

\maketitle

\abstract{Reheating or entropy production on the three-brane 
from a decaying bulk scalar field is studied in the brane-world picture of
the Universe.  It is
shown that a significant amount of
dark radiation is produced in this process unless only the modes which
satisfy a specific relation 
are excited, so that subsequent entropy
production within the brane is required in general
before primordial nucleosynthesis. 
}

\pacs{PACS Numbers:  98.80.Cq;04.50.+h;11.25.Mj;12.10.-g  ~~~~~~~~~OU-TAP-158}

\maketitle

%\newpage
%
%\def\theequation{\arabic{section}.\arabic{equation}}

        %%%%%%%%        Contents starts here  %%%%%%%%%%%%%%
\newcommand{\dw}{{\rm DW}}
\newcommand{\cw}{{\rm CW}}
\newcommand{\ml}{{\rm ML}}
\newcommand{\lt}{\tilde{\lambda}}
\newcommand{\lh}{\hat{\lambda}}
\newcommand{\phidot}{\dot{\phi}}
\newcommand{\phiw}{\phi_{,w}}
\newcommand{\sqg}{\sqrt{-g}}
\newcommand{\phicl}{\phi_{cl}}
\newcommand{\adot}{\dot{a}}
\newcommand{\phat}{\hat{\phi}}
\newcommand{\ahat}{\hat{a}}
\newcommand{\hhat}{\hat{h}}
\newcommand{\phihat}{\hat{\phi}}
\newcommand{\Nhat}{\hat{N}}
\newcommand{\hth}{h_{th}}
\newcommand{\hbh}{h_{bh}}
\newcommand{\gsim}{\gtrsim}
\newcommand{\lsim}{\lesssim}
\newcommand{\bfx}{{\bf x}}
\newcommand{\bfy}{{\bf y}}
\newcommand{\bfr}{{\bf r}}
\newcommand{\bfk}{{\bf k}}
\newcommand{\bkp}{{\bf k'}}
\newcommand{\order}{{\cal O}}
\newcommand{\beq}{\begin{equation}}
\newcommand{\eeq}{\end{equation}}
\newcommand{\beqa}{\begin{eqnarray}}
\newcommand{\eeqa}{\end{eqnarray}}
\newcommand{\bea}{\begin{eqnarray}}
\newcommand{\eea}{\end{eqnarray}}
\newcommand{\mpl}{M_{Pl}}
\newcommand{\lmk}{\left(}
\newcommand{\rmk}{\right)}
\newcommand{\lkk}{\left[}
\newcommand{\rkk}{\right]}
\newcommand{\lnk}{\left\{}
\newcommand{\rnk}{\right\}}
\newcommand{\call}{{\cal L}}
\newcommand{\calr}{{\cal R}}
\newcommand{\half}{\frac{1}{2}}
\newcommand{\kc}{k_c}
\newcommand{\lambdac}{\lambda_c}
\newcommand{\bkc}{\beta\kappa\chi}
\newcommand{\gkc}{\gamma\kappa\chi}
\newcommand{\gbkc}{(\gamma-\beta)\kappa\chi}
\newcommand{\dchi}{\delta\chi}
\newcommand{\dphi}{\delta\phi}
\newcommand{\dOmega}{\delta\Omega}
\newcommand{\Phibd}{\Phi_{\rm BD}}
\newcommand{\echi}{\epsilon_\chi}
\newcommand{\ephi}{\epsilon_\phi}
\newcommand{\Phihat}{\hat{\Phi}}
\newcommand{\Psihat}{\hat{\Psi}}
\newcommand{\that}{\hat{t}}
\newcommand{\Hhat}{\hat{H}}
\newcommand{\zk}{z_k}
\newcommand{\msolar}{M_\odot}
\newcommand{\mbh}{M_{\rm BH}}
\newcommand{\bh}{{\rm BH}}
\newcommand{\calf}{{\cal F}}
\newcommand{\gtilde}{{\tilde g}}
\newcommand{\lcdm}{$\Lambda$CDM}
\newcommand{\chione}{{\chi_1}}
\newcommand{\chionem}{\chi_{1m}}
\newcommand{\chitwo}{{\chi_2}}
\newcommand{\phione}{{\phi_1}}
\newcommand{\phionec}{\phi_{1c}}
\newcommand{\phitwo}{{\phi_2}}
\newcommand{\phitwom}{\phi_{2m}}
\newcommand{\phitwoc}{\phi_{2c}}
\newcommand{\Vone}{V_1}
\newcommand{\Vtwo}{V_2}
\newcommand{\vone}{v_1}
\newcommand{\vtwo}{v_2}
\newcommand{\mone}{m_1}
\newcommand{\mtwo}{m_2}
\newcommand{\lone}{\lambda_1}
\newcommand{\ltwo}{\lambda_2}
\newcommand{\gone}{g_1}
\newcommand{\gtwo}{g_2}
\newcommand{\alphaone}{\alpha_1}
\newcommand{\alphatwo}{\alpha_2}
\newcommand{\mg}{M_5}
\newcommand{\my}{M_4}
\newcommand{\kg}{\kappa_5}
\newcommand{\ky}{\kappa_4}
\newcommand{\lz}{\lambda_0}
\newcommand{\rhot}{\rho_{\rm tot}}
\newcommand{\gb}{\Gamma_B}
\newcommand{\gd}{\Gamma_D}
\newcommand{\lan}{\lambda_n}
\newcommand{\mn}{m_n}

%\baselineskip 0.8cm
%\newpage

Any new theory of gravity and/or high energy physics must pass a
number of cosmological tests, among which is implementation of
cosmological inflation \cite{oriinf,inf}.  Successful inflation must
fulfill three requirements, namely, 
sufficiently long quasi-exponential expansion driven by vacuum-like
energy density such as a potential energy of a scalar field, 
termination of accelerated expansion associated with entropy production
or reheating to set the initial state of the  classical hot Big Bang
cosmology well before the primordial nucleosynthesis  \cite{ns}, and
generation of primordial fluctuations with desired amplitude and
spectrum \cite{fluc}.  
It is much more difficult to achieve the second element than the first 
in general, and the third one typically requires fine tuning of
model parameters.

In this paper we consider reheating after inflation in the brane world
picture of the Universe \cite{RS,RS2}.  
In this scenario, our Universe is described on the four-dimensional
boundary (three-brane) of $Z_2$-symmetric five-dimensional spacetime
with a negative cosmological constant $\Lambda_5\equiv -6k^2$, where 
$k$ is a positive constant.
This situation not only takes into account the spirit of 
Horava-Witten theory \cite{horava,lukas}, 
but also recovers the Einstein gravity around the
brane with positive tension \cite{RS2,tanaka,SMS}.  Much work has
been done on brane-world cosmology \cite{cosm,bki,mukoyama} 
including inflationary brane solutions \cite{binf,bas,bbinf,HS}.

We assume the five-dimensional Einstein gravity with a negative
cosmological constant $\Lambda_5$ and a three-brane 
at the fifth coordinate $w=0$ about which the spacetime is $Z_2$
symmetric.  
We write the metric near the
brane in the following form in terms of the Gaussian normal coordinate.
\bea
ds_5^2&=&g_{AB}dx^Adx^B
=-N^2(t,w)dt^2+Q^2(t,w)a^2(t)\left
( {dx^2+dy^2+dz^2} \right)+dw^2\equiv q_{\mu\nu}dx^\mu dx^\nu+dw^2, \label{senso}
\eea
where capital Latin indices run 0,1,2,3, and 5 while Greek indices from 0
to 3.  We take $N=Q=1$ on the brane $w=0$.  One can write down functional forms
of $N(t,w)$ and $Q(t,w)$ explicitly   as
\beqa
Q^2(t,w)&=&\cosh (2kw)+{\textstyle{1 \over 2}}k^{-2}H^2\left( {\cosh
(2kw)-1} \right)- \sqrt {1+k^{-2}H^2+Ca^{-4}}\sinh (2k|w|),  \nonumber \\ 
N^2 (t,w)&=&Q^{-2}(t,w)\lkk\cosh (2kw)+{\textstyle{1 \over 2}}k^{-2}\left( {H^2+\dot
H} \right)\left( {\cosh (2kw)-1} \right)- {{1+{\textstyle{1 \over
2}}k^{-2}\left( {2H^2+\dot H} \right)} \over {\sqrt
{1+k^{-2}H^2+Ca^{-4}}}}\sinh (2k|w|)\rkk^2, \label{metric} 
\eeqa
 in the case the
bulk is in a vacuum state with a negative cosmological constant 
$\Lambda_5$, where $C$ is an integration
constant \cite{mukoyama}.  
In this solution
the induced metric on the brane is nothing but the spatially flat
Robertson-Walker metric with the scale factor
$a(t)$.

The evolution equation on the brane in this case is given by
\beqa
H^2&=&\lmk\frac{\dot{a}}{a}\rmk^2
={{\kappa _5^4\sigma } \over {18}}\rhot +{{\Lambda _4} \over
3}+{{\kappa _5^4} \over {36}}\rhot ^2-{{k^2C} \over {a^4}}, \label{feq} \\ 
\Lambda_4&\equiv&{1 \over 2}\left( {\Lambda _5+{{\kappa _5^2} \over
6}\sigma ^2} \right), \label{Friedmann}
\eeqa
where $\kg^2$ is the five dimensional gravitational constant related with
the five dimensional reduced Planck scale, $\mg$, by $\kg^2=\mg^{-3}$. Here
$\sigma$ is the brane tension, $\rhot$ is the total energy density on
the brane, and the last term of (\ref{feq}) represents the so-called
dark radiation with $C$ being an integration constant  \cite{SMS,bki,mukoyama}.
In order to recover the standard Friedmann equation with a vanishing
cosmological constant at low energy scales, we require 
$\sigma=6k/\kg^2$ and $\ky^2=\kg^4\sigma/6=\kg^2k$, where 
$\ky^2$ is the four
dimensional gravitational constant related with the four dimensional 
reduced Planck scale, $\my$, as $\ky^2=\my^{-2}$.  
We therefore find 
$\my^2=\mg^3/k$.  That is, if we take $k=\my$, all the fundamental
scales in the theory take the same value, $k=\my=\mg$.  
Note that $k$ also sets the scale above which the
nonstandard term quadratic in $\rhot$ is effective in (\ref{feq}).
We assume that $k$ is much larger than the scale of inflation so that
such quadratic corrections are negligible.

We consider the case inflation is driven by a bulk scalar field $\phi$
with a five-dimensional potential $V[\phi]$ \cite{bbinf,HS} and study
the evolution of $\phi$ after brane inflation, because reheating is expected to
proceed in the same way as in four dimensional theory if the inflaton lives
only on the brane \cite{bas}.  Since $\phi$
is homogenized in three space as a result of inflation, it
depends only on $t$ and $w$.  We consider a situation
 that $\phi$ rapidly oscillates 
around $\phi=0$  and assume that $V[\phi]$ is expressed as
$V[\phi]=m^2\phi^2/2$.
Then the Klein-Gordon equation reads
\beq
 \Box_5\phi(w,t)-V'[\phi(w,t)]=\frac{1}{\sqg}\partial_t\lmk\sqg g^{00}\phidot\rmk
 +\frac{1}{\sqg}\partial_w\lmk\sqg\phiw\rmk-V'[\phi]=0, \label{kg}
\eeq
where a dot denotes time differentiation.
In order to express energy release of $\phi$ we
introduce the  following dissipation term phenomenologically in  (\ref{kg}).
\beq
 \Box_5\phi(w,t)-V'[\phi(w,t)]=\frac{\gd}{2k}\delta(w)\frac{1}{N}\phidot
  +\gb\frac{1}{N}\phidot. \label{kgv}
\eeq
Here $\gd$ and $\gb$ represent energy release to the brane and to the
entire space, respectively.  The  denominator in the right-hand-side  is
introduced on dimensional grounds.

From (\ref{kg}) and (\ref{kgv}) together with the $Z_2$ symmetry,
we find 
\beq
\phiw^+=-\phiw^-=\frac{\gd}{4k}\phidot(0,t),  \label{phiw}
\eeq
where superscripts $+$ and $-$ imply values at $w\longrightarrow
+0$ and $-0$, respectively.  

The divergence of the energy-momentum tensor of
the scalar field,
\beq
  T_{MN}^{(\phi )}=
\phi _{,M}\phi _{,N}-g_{MN}\left( {{\textstyle{1 \over 2}}g^{PQ}
\phi_{,P}\phi _{,Q}
+V[\phi]} \right),
\eeq
reads,
\beq
 T^{(\phi)C}_{~~A;C}=\lnk \Box_5\phi(w,t)-V'[\phi(w,t)]\rnk \phi_{,A}
 =\lkk \frac{\gd}{2k}\delta(w)\frac{1}{N}\phidot
  +\gb\frac{1}{N}\phidot\rkk \phi_{,A}.  \label{divergence}
\eeq
Integrating $A=0$ component of 
(\ref{divergence}) from $w=-\epsilon$ to $w=+\epsilon$ near
the brane, we find (\ref{phiw}) from the zeroth order in $\epsilon$ and
\beq
 \frac{\partial\rho_\phi(0,t)}{\partial t}
=-(3H+\gb)\phidot^2(0,t)-J_\phi(0,t),    \label{rhophieq} 
\eeq
with
\beq
\rho_\phi \equiv
\frac{1}{2}\phidot^2+V[\phi],
~~~J_\phi \equiv -\frac{\phidot}{\sqg}\partial_w\lmk\sqg\phiw\rmk,
\eeq
from the terms proportional to $\epsilon$.
  Thus the energy dissipated by the $\gd$ term
on the brane
is entirely compensated by the energy flow onto the brane.

Next we study how the energy released from $\phi$ affects evolution of
our brane Universe by analyzing gravitational field equations
\cite{SMS,HS}.  In the present situation the total energy momentum tensor
including the contribution of bulk cosmological constant reads
\beq
T_{MN}=-\kappa _5^{-2}\Lambda _5g_{MN}+T_{MN}^{(\phi )}+S_{MN}\delta
(w), 
\eeq
where $S_{MN}$ is the stress tensor on the brane.  Its nonvanishing
components can be further decomposed as
\beq
  S_{\mu\nu}=-\sigma q_{\mu\nu}+\tau_{\mu\nu}.
\eeq
Here $\tau_{\mu\nu}$ represents energy momentum tensor of the radiation
fields produced by decay of $\phi$ and it is of the form
$\tau^{\mu}_{\nu}={\rm diag}(-\rho_r,p_r,p_r,p_r)$ with $p_r=\rho_r/3$.

In terms of the unit vector $n_M=(0,0,0,0,1)$ normal to the brane,
the extrinsic curvature of a $w=$constant hypersurface is given by
$K_{MN}=q_M^Pq_N^Qn_{Q;P}$ with $q_{MN}=g_{MN}-n_Mn_N$.  Then from 
the Codazzi equation and the five dimensional Einstein equation, we find
\beq
D_\nu K_\mu ^\nu -D_\mu K=\kappa _5^2T_{MN}n^Nq_\mu ^M
=\kappa _5^2T_{\mu w}=\kappa _5^2\dot{\phi}\phi _{,w}\delta _\mu ^0, \label{k1}
\eeq
where $D_\nu$ stands for the four dimensional covariant derivative with respect to the
metric $q_{\mu\nu}$.
The above equation reads 
\beq
D_\nu K_0^{\nu+} -D_0K^+ =\kappa _5^2{\gd  \over {4k}}\dot \phi ^2(0,t), \label{k1a}
\eeq 
near the brane, $w\longrightarrow +0$.

From the junction condition and the $Z_2$-symmetry, on the other hand,
we find
\beq
K_{\mu \nu}^+=-{{\kappa _5^2} \over 2}
\left( {S_{\mu \nu }-{1 \over 3}q_{\mu \nu }S} \right), 
\eeq
therefore
\beq
D_\nu K_\mu ^{\nu+} -D_\mu K^+
=-{{\kappa _5^2} \over 2}D_\nu S_\mu ^\nu 
=-{{\kappa _5^2} \over 2}D_\nu \tau _\mu ^\nu. \label{k2}
\eeq
Combining (\ref{k1a}) and (\ref{k2}), we obtain
\beq
D_\nu \tau _\mu ^\nu =-{\gd  \over 2k}\dot \phi ^2\delta _\mu ^0,
\eeq
namely,
\beq
{{\partial \rho_r } \over {\partial t}}=-3H(\rho_r +p_r)+{\gd  \over 2k}\dot
\phi ^2=-4H\rho_r+{\gd  \over 2k}\dot\phi ^2,
\eeq
on the brane.  
Thus we find 
only the dissipation term proportional to $\gd$ with the delta
function is effective to reheat the brane. 
This equation has the same
form as the reheating in perturbation theory after conventional  inflation
in four dimensional theory \cite{prt}.

On the other hand, the four dimensional Einstein tensor,
 $G^{(4)\nu}_{~\mu}$, satisfies the following equality on the brane \cite{HS},
\beq
G^{(4)\nu}_{~\mu} =
\kappa _4^2\left( {T^{(s)\nu}_{~\mu}  +\tau _\mu ^\nu } \right)
+\kappa _5^4\pi _\mu ^\nu -E_\mu ^\nu,
\eeq
with
\beq
T^{(s)\nu}_{~\mu}  \equiv{1 \over {6k}}
\left[ {4q^{\nu \zeta }\phi _{,\mu }\phi _{,\varsigma }
+\left( {{3 \over 2}\phi _{,w}^2
-{5 \over 2}q^{\xi \zeta }\phi _{,\xi }\phi _{,\varsigma }
-\frac{3}{2}m^2\phi^2} \right)q_\mu ^\nu } \right].
\eeq 
Here $\pi^{\nu}_\mu$ represents terms quadratic in $\tau _\alpha ^\beta$
which are higher order in $\rho_r/(k\my)^2$ and are consistently neglected in
our analysis.  $E_\mu ^\nu\equiv C^{w\nu}_{\mu w}$  
is a component of the five dimensional Wyel tensor $C^{MN}_{PQ}$, which
is the origin of the dark radiation \cite{mukoyama}.

Now we write down the four dimensional Bianchi identity,
\beq
D_\nu G^{(4)\nu}_{~\mu}  =0=
\kappa _4^2\left( {D_\nu T^{(s)\nu}_{~\mu} 
+D_\nu \tau _\mu ^\nu } \right)-D_\nu E_\mu ^\nu, \label{bianchi} 
\eeq
to yield
\beq
D_\nu E_0^\nu =-{{\kappa _4^2} \over {2k}}
{\partial  \over {\partial t}}\left( {{1 \over 2}\dot \phi ^2
-{1 \over 2}\phi _{,w}^2+\frac{1}{2}m^2\phi^2} \right) - \frac{2\ky^2H}{k}\phidot^2
-{{\kappa _4^2} \over 2k}\gd \dot \phi ^2. \label{19}
\eeq
Note that $\phi_{,w}^2=\gd^2\phidot^2/(16k^2)$ on the brane and is
negligibly small compared with $\phidot^2$, 
since we expect the dissipation rate $\gd$ is at most comparable to the
scale of inflation and hence much
smaller than the scale $k$.
With this approximation the quantity in the parenthesis in (\ref{19})
may be replaced by $\rho_\phi(0,t)$. 
We also find $D_\nu E_0^\nu =\partial _0E_0^0+4HE_0^0$ because
$E^{\nu}_\mu$ is traceless.

As a result we obtain the following set of evolution equations in the
brane universe $w=0$ in terms of $\varphi(t)\equiv\phi(0,t)/\sqrt{2k}$.
\beqa
H^2&=&\left( {{{\dot a} \over a}} \right)^2
={{\kappa _4^2} \over 3}\left( \rho _\varphi+\rho_r  +\varepsilon \right),
~~~\rho _\varphi \equiv  {{1 \over 2}\dot \varphi ^2+
\frac{1}{2}m^2\varphi^2=\frac{\rho_\phi}{2k} },
~~~\varepsilon\equiv \frac{E^0_0}{\kappa_4^2},
 \label{22} \\
\frac{\partial\rho_\varphi}{\partial t}
&=&-(3H+\gb)\dot\varphi^2-J_\varphi, ~~~J_\varphi \equiv
\frac{J_\phi}{2k}, \label{23} \\
{{\partial \rho_r } \over {\partial t}}
&=&-4H\rho_r+\gd\dot \varphi^2, \label{req} \\
{{\partial \varepsilon} \over {\partial t}}&=&
-4H\varepsilon -(H+\gd-\gb)\dot{\varphi}^2+J_\varphi,
\label{eeq}
\eeqa
where we have used (\ref{rhophieq}).

Here $J_\varphi$ represents energy flow of the scalar field into the
bulk and we cannot calculate it unless we solve the field
equation in the bulk.  It is, however, formidable to do so in the
realistic situation that the bulk metric is influenced by the bulk
scalar field which would be the case in brane inflation driven by a bulk
scalar field.  In this case the metric would be even more complicated
than (\ref{metric}).  Hence in order to obtain some insights on the form
of $J_\varphi$ we must employ several approximations and find a bulk
scalar solution.  

Let us assume the bulk metric is governed by
$\Lambda_5$ and further neglect terms suppressed by $k^{-1}$ and
$Ca^{-4}$ in (\ref{metric}).  Then the metric reads 
\beq
ds^2_5=-e^{-2k|w|}dt^2+e^{-2k|w|}a^2(t)\left
( {dx^2+dy^2+dz^2} \right)+dw^2, \label{met}
\eeq
 and field equations (\ref{kg}) and
(\ref{kgv}) become more tractable.  In the time scale of the field
oscillation the dissipation terms are unimportant because they are
perturbatively small quantities by assumption.  
So let us  estimate $J_\varphi$ by solving
(\ref{kg}) instead of (\ref{kgv}) first.
The solution of (\ref{kg}) under the metric (\ref{met}) is easily found 
in the same way as Goldberger and Wise \cite{GW} as
\beq
\phi (t,w)=\sum\limits_n {c_nT_n(t)Y_n(w)}+H.C.
\eeq
with
\beqa
T_n(t)&\cong& a^{-{\textstyle{3 \over 2}}}(t)e^{-im _nt}, \\
Y_n(w)&=&e^{2k|w|}\left[ {J_\nu ^{}
\left( {{\textstyle{{m _n} \over k}}e^{k|w|}} \right)
+b_nN_\nu ^{}\left( {{\textstyle{{m _n} \over k}}e^{k|w|}} \right)} \right],~~~
\nu =2\sqrt {1+{{m^2} \over {4k^2}}}\cong 2+{{m^2} \over {4k^2}},
\eeqa
under the assumption that the field oscillates rapidly in cosmic
expansion time scale.  Here $m_n$ is a separation constant which may
take continuous values in the present case with a single brane, and
$b_n$ is a constant determined by the boundary condition, $\phi_{,w}=0$, at $w=0$
 as
\beq
b_n=\lkk 2 J_\nu\lmk\frac{m_n}{k}\rmk
+\frac{m_n}{k}J'_\nu\lmk\frac{m_n}{k}\rmk\rkk
\lkk 2 N_\nu\lmk\frac{m_n}{k}\rmk
+\frac{m_n}{k}N'_\nu\lmk\frac{m_n}{k}\rmk\rkk^{-1}.
\eeq
If we incorporate the effect of dissipation on the boundary condition,
(\ref{phiw}), it is modified to
\beq
b_n\cong\lkk \lmk 2+\frac{im_n\gd}{2k^2}\rmk J_\nu\lmk\frac{m_n}{k}\rmk
+\frac{m_n}{k}J'_\nu\lmk\frac{m_n}{k}\rmk\rkk
\lkk\lmk 2+\frac{im_n\gd}{2k^2}\rmk N_\nu\lmk\frac{m_n}{k}\rmk
+\frac{m_n}{k}N'_\nu\lmk\frac{m_n}{k}\rmk\rkk^{-1},
\eeq
where use has been made of $\dot{T_n}(t)\cong -im_nT_n(t)$.

Hereafter let us  assume only
a single oscillation mode exists.  This is a natural assumption because
any higher mode is expected to decay earlier and the final stage of
reheating would be dominated by the lowest mode that has been excited.
Then we find
\beq
 J_\varphi=(m_n^2-m^2)\varphi\dot\varphi. \label{jvarphi}
\eeq

\if
Then we can calculate evolution of the dark radiation if that of
$\varphi(t)\equiv \phi(0,t)/\sqrt{2k}$ is specified.  Since we are
interested in the regime $\varphi(t)$ oscillates rapidly, let us
parameterize its evolution as
\beq
  \varphi(t) =
\varphi_i\lmk\frac{a(t)}{a(t_i)}\rmk^{-3/2}e^{-i\lan (t-t_i)},
~~~~~\lan\equiv\mn-\frac{i}{2}\gb,
\eeq
with $\mn \gg H$ and $\gb$ being positive constants.  That is, we assume only
a single oscillation mode exists.  This is a natural assumption because
any higher mode is expected to decay earlier and the final stage of
reheating would be dominated by the lowest mode that has been excited.
The imaginary part of $\lan$, $\gb$, which  is presumably much smaller than
its real part, is related to, but in general different from $\gd$.  Its
magnitude depends on both the underlying theory and the far out boundary
condition.  Here we treat it as a free parameter in order to investigate
various situations.  Then the evolution equation of 
$\varepsilon\equiv \kappa_4^{-2}E^0_0$ is given by
\beq
  \frac{\partial \varepsilon}{\partial t}=-4H\varepsilon 
 -(\gd+H-\gb)\dot{\varphi}^2-(m^2-\mn^2)\varphi\dot{\varphi}.
\eeq
\fi
Since $\varphi$ oscillates rapidly in the expansion time scale let us
average the right-hand-side of evolution equations
(\ref{23})--(\ref{eeq}) over an oscillation period.
Using $\overline{\dot{\varphi}^2}(t)=\mn^2\overline{\varphi^2}(t)$ and
$\overline{\varphi\dot{\varphi}}(t)=-(3H+\gb)\overline{\varphi^2}(t)/2$,
we obtain the following set of evolution equations in the
brane universe $w=0$ where a bar denotes average over the oscillation period.
\beqa
H^2&=&\left( {{{\dot a} \over a}} \right)^2
\cong {{\kappa _4^2} \over 3}\left( \rho _\varphi+\rho_r  +\varepsilon \right),
\\
\frac{\partial\rho_\varphi}{\partial t}&=&
-\frac{1}{2}(3H+\gb)(m^2+m_n^2)\overline{\varphi^2}, \\
%\equiv-(3H+\gb)(1+\omega_\varphi)\rho_\varphi, \label{23b} \\
{{\partial \rho_r } \over {\partial t}}
&=&-4H\rho_r+\gd\mn^2\overline{\varphi^2}, \label{reqb} \\
{{\partial \varepsilon} \over {\partial t}}&=&
-4H\varepsilon -(\gd+H-\gb)\mn^2\overline{\varphi^2} 
+\frac{1}{2}(3H+\gb)(m^2-\mn^2)\overline{\varphi^2},
\label{eeqb} 
\eeqa
with
\beqa
\overline{\varphi^2}(t)\equiv \overline{\varphi^2_i}
\lmk\frac{a(t)}{a(t_i)}\rmk^{-3}e^{-\gb (t-t_i)}. \label{varphi}
\eeqa
\if
Although the above equations depend solely on quantities on the brane
and they resemble those in the case of perturbative reheating in the ordinary
inflation to this end \cite{prt}, except of course for those involving
$\varepsilon$, we cannot solve them as they are.  This is because the
effective equation of state of $\varphi$, which is characterized by
$\omega_\varphi$ defined in (\ref{23}), is in general different from
that of dust even when the mass term dominates the potential,
 due to the presence of the extra dimension or Kalza-Klein modes
\cite{GW}.  In fact, 
$\omega_\varphi$ depends on the model of inflation as well as on the initial condition.
Furthermore in general it changes in time.  
\fi

We assume that both $\rho_r$
and $\varepsilon$ are vanishingly small at the end of inflation when we set the
initial condition, because they
rapidly redshift during inflation and very little creation is expected
in that period \cite{HS}.  We identify the reheating epoch with the time
when $\rho_\varphi$ becomes smaller than $\rho_r$.  From (\ref{varphi})
this occurs at $t\simeq H^{-1}\simeq \gb^{-1}$, so that the dominant
creation takes place during $H\gtrsim \gb$.  Here we qualitatively analyze 
the system for various values of $\mn$.

\begin{description}
\item[Case A: {\boldmath $\mn \geq m$}] ~\\
In this case, the creation terms of dark radiation, 
$-(\gd+H-\gb)\mn^2\overline{\varphi^2} 
+\frac{1}{2}(3H+\gb)(m^2-\mn^2)\overline{\varphi^2}$ are negative and
their magnitude is larger than the creation term of radiation,
$\gd\mn^2\overline{\varphi^2}$ during the important period $H \gtrsim \gb$.
Hence more dark radiation is created than ordinary
radiation in magnitude.  Then $\rho_r$ and $\varepsilon$ tend to cancel
each other and the higher-order terms of
the Friedmann equation, which we have neglected so far,
would play an important role in the subsequent evolution of the three
brane.  This  means that
we do not recover standard cosmology on the brane after inflation.
\item[Case B: {\boldmath $\mn \ll m$}] ~\\
This possibility is not excluded in the brane-world scenario with
	   non-factorizable geometry unlike in the case of ordinary
	   Kalza-Klein compactification with factorizable metric
	   \cite{GW}.
In this case the last term of (\ref{eeqb}) is dominant and we find more
	   dark radiation than ordinary radiation unless $\gb$ is
	   extremely small with $\gb/\gd < \mn^2/m^2 \ll 1$.
\item[Case C: {\boldmath $\mn \lesssim m$}] ~\\
This is the most delicate case and the final amount of dark radiation can
	   be either positive or negative depending on the details of the model
	   parameters.  But in order to achieve successful primordial 
	   nucleosynthesis, the amount of extra radiation-like matter is
	   severely constrained \cite{ns}.  In order to have sufficiently small
	   $\varepsilon$ compared with $\rho_r$ after reheating without
	   resorting to subsequent entropy production within the brane,
the magnitude of creation terms of $\varepsilon$ should be vanishingly
	   small at the reheating epoch $H\simeq \gb$.  That is, only
	   the mode that satisfies the
	   inequality 
\beq
  \left| 2\gb(m^2-\mn^2) -\gd\mn^2\right| \ll \gd\mn^2,  \label{cond}
\eeq
should be present.
\end{description}

We thus find a specific relation (\ref{cond}) should be satisfied for
the graceful exit of brane inflation  driven by a bulk scalar field $\phi$.
In case it is not satisfied, we must introduce some
other mechanisms of entropy production 
within the brane before primordial nucleosynthesis
which imposes stringent constraints on 
 the exotic energy density and the expansion law at that era \cite{ns}.
One way is to assume that $\phi$ predominantly decays into massive
particles whose energy density redshifts less rapidly and dominates
over $\varepsilon$ soon.  If these particles decay into radiation before nucleosynthesis 
creating an appropriate amount of baryon asymmetry, we may recover the
standard cosmology.  Another way is to consider second inflation which
is driven by a scalar field confined on the brane \cite{bas} to dilute $\varepsilon$.

In summary we have studied entropy production on the three-brane from a
decaying bulk scalar field $\phi$ by introducing  dissipation terms to
its equation of motion phenomenologically.  We have  shown that the so-called dark
radiation is significantly  produced at the same time unless the
inequality (\ref{cond}) is satisfied.  Although we have analyzed only 
the case with a  specific form of the dissipation, we expect our
conclusion is generic and applicable to other forms of dissipation, too,
because it is essentially an outcome of the four
dimensional Bianchi identity (\ref{bianchi}).
We therefore conclude that in the brane-world picture of the Universe,
it is likely that 
the dominant part of the entropy we observe today originates within the
brane rather than in the bulk.

\vskip 0.4cm
%\acknowledgements{
The authors are grateful to M.\ Sasaki and T.\ Tanaka for useful
communications.  The work of JY was partially supported by the Monbukagakusho
Grant-in-Aid, Priority Area ``Supersymmetry and Unified Theory of
Elementary Particles''(\#707) and the Monbukagakusho Grant-in-Aid for
Scientific Research Nos.\ 11740146 and 13640285. %} 

\end{document}